# Giant Nonreciprocity of Surface Acoustic Waves enabled by the Magnetoelastic Interaction

**Authors**

Piyush J. Shah[1], Derek A. Bas[1], Ivan Lisenkov[2], Alexei Matyushov[3, 4], Nianxiang Sun[3], Michael R. Page[1*]

**Affiliations**

[1] Materials and Manufacturing Directorate, Air Force Research Laboratory, Wright-Patterson Air Force Base, Ohio 45433, USA
[2] Independent Researcher, Newton Upper Falls, MA 02464, USA
[3] Department of Electrical and Computer Engineering, Northeastern University, Boston, MA 02115, USA
[4] Department of Physics, Northeastern University, Boston, MA 02115, USA

[*]Email: michael.page.16@us.af.mil

**Abstract**

Nonreciprocity, the defining characteristic of isolators, circulators and a wealth of other applications in radio/microwave communications technologies, is in general difficult to achieve as most physical systems incorporate symmetries that prevent the effect. In particular, acoustic waves are an important medium for information transport, but they are inherently symmetric in time. In this work, we report giant nonreciprocity in the transmission of surface acoustic waves (SAWs) on lithium niobate substrate coated with ferromagnet/insulator/ferromagnet ($FeGaB/Al_2O_3/FeGaB$) multilayer structure. We exploit this novel structure with a unique asymmetric band diagram, and expand on magnetoelastic coupling theory to show how the magnetic bands couple with acoustic waves only in a single direction. We measure 48.4 dB (ratio of 1:100,000) isolation which outperforms current state of the art microwave isolator devices in a novel acoustic wave system that facilitates unprecedented size, weight, and power reduction. Additionally, these results offer a promising platform to study nonreciprocal SAW devices.

**Introduction**

Nonreciprocal microwave transmission devices such as isolators and circulators have an important role in the front-end of most RF systems and test and measurement equipment. These devices allow RF propagation in one direction and block propagation in the opposite direction. From an application point of view, the ideal attributes of an isolator device should include low insertion loss (allow full transmission from port 1 to port 2) and high isolation (block transmission from port 2 to port 1). This type of device can be thought of as a diode for RF energy. Current state of the art isolators and circulators utilize a transversely magnetized ferrite junction to direct the incoming microwave energy and thus allowing travel in the direction of the magnetizing field. In 1971, Lewis proposed an alternative form of acoustic isolator device concept using a layered SAW delay line with ZnO/YIG on GGG substrate [1]. While acoustic isolator device concepts have largely been ignored for decades, these concepts are the subject of very recent theoretical investigations generating significant interest in the scientific community [2, 3, 4]. In general, nonreciprocal propagation of SAWs is nontrivial to achieve and has been observed in nonmagnetic metal (aluminum) and some semiconductor heterostructures [5]. However, the nonreciprocity magnitude is not sufficient for real-world application relevance. On the other hand, spin wave nonreciprocity has been an active area of research interest in the community resulting in numerous reports in the last decade or so [6, 7, 8, 9, 10]. The theoretical framework that explains spin wave nonreciprocity is either based on frequency displacement in the ferromagnetic layer or is based on interband magnonic transitions in a system with lack of time-reversal symmetry [6, 9, 11].

Recently, the community has been investigating device physics utilizing magnetoelastic interactions of spin and acoustic waves. This is based on traveling SAWs coupling into the magnetostrictive ferromagnetic thin film in the SAW propagation path. The most common materials system studied on this subject is Ni on lithium niobate ($LiNbO_3$) which has been shown to have reciprocal transmission behavior due to polycrystallinity of the Ni film. This leads to larger Gilbert damping coefficient resulting in wider line widths in the magnetization response [12]. Several device concepts such as magnetically tunable phase shifters and resonators were reported in the 1970s utilizing magnetoelastic interactions. The recent resurgence of study in magnetoelastic interactions utilizing SAWs is being termed as acoustically driven ferromagnetic resonance (ADFMR) [13, 14, 15].

SAW-based frequency filters, delay lines and sensors are mature technologies and have several applications in the RF frequency (low MHz up to 2.5 GHz) regime. Ultra-low loss, temperature compensated SAW filters are essential elements in military and consumer communication devices such as cell phones and tablets. Acoustic transmission is advantageous because the propagation speeds and wavelengths are typically several orders of magnitude lower than for electromagnetic waves and therefore scaling down is easily achieved.

In this work, we demonstrate giant nonreciprocity of surface acoustic waves through the magnetoelastic interaction and operation of a magnetoelastic magnetic field-dependent microwave isolator. We utilize a novel ferromagnetic/dielectric heterostructure in a traditional SAW delay line filter geometry which achieves record high RF isolation and nonreciprocal behavior, thereby opening a new avenue to explore next generation size, weight, and power-friendly microwave isolator and circulator devices. A device schematic is shown in Fig. 1.

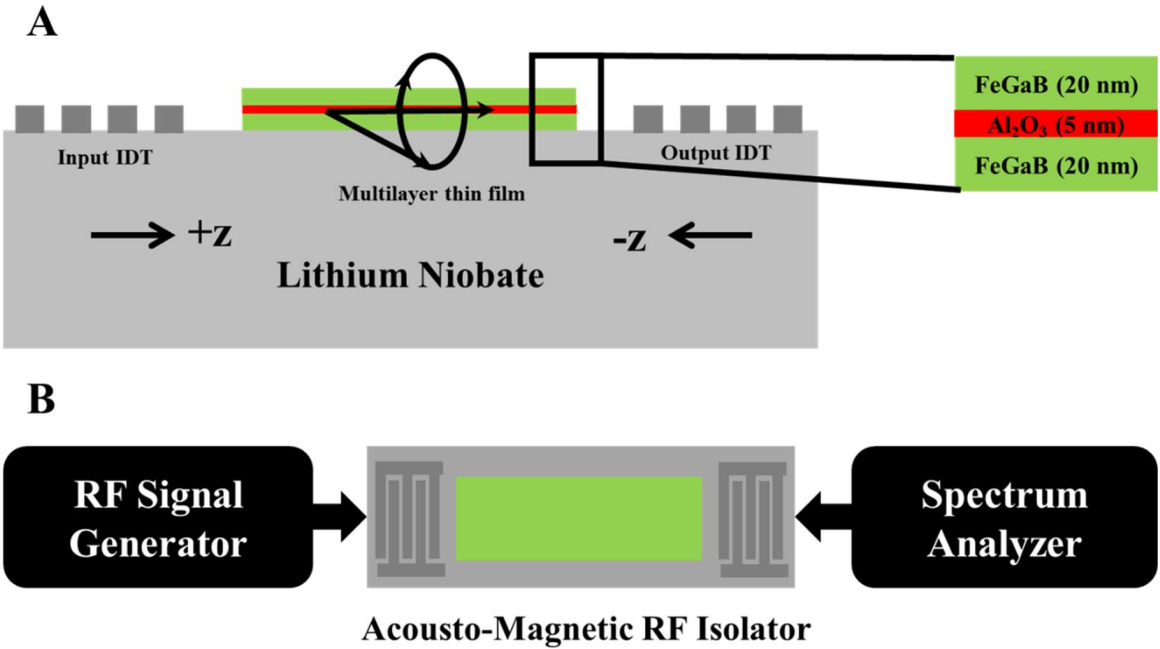

**Fig. 1. Schematic of device design and experimental setup for measurement.** (**A**) Schematic cross-section representation of the magnetoelastic isolator device. The geometry of the device is similar to the standard ADFMR device. The multilayer thin film stack in this study is FeGaB/Al$_2$O$_3$/FeGaB thin film. $L$ is the thickness of magnetic film and $d_s$ is the thickness of dielectric (spacer) film. (**B**) Schematic top view representation of the isolator device including measurement setup.

**Results**

Fig. 2 shows the conventional angle-dependent magnetic field sweep versus transmission magnitude plot from the ADFMR device driven at 1435 MHz. Resonance field, linewidth, and contrast are quantitatively similar to that observed in a single-layer FeGaB device [16]. Here, the growth field $H_G$ was applied at 60° relative to $+z$ (see inset). In all prior reports of ADFMR in Ni, broad absorption resonances occurred in all four quadrants with even and odd symmetry. Notable distinctions of our measurement include: (1) breaking even symmetry, with lobes occurring only in quadrants II and IV; (2) extremely narrow line width, indicating low damping and therefore high frequency selectivity; (3) acoustic wave absorption much larger than any previously reported results; and (4) high nonreciprocity in opposite SAW travel directions.

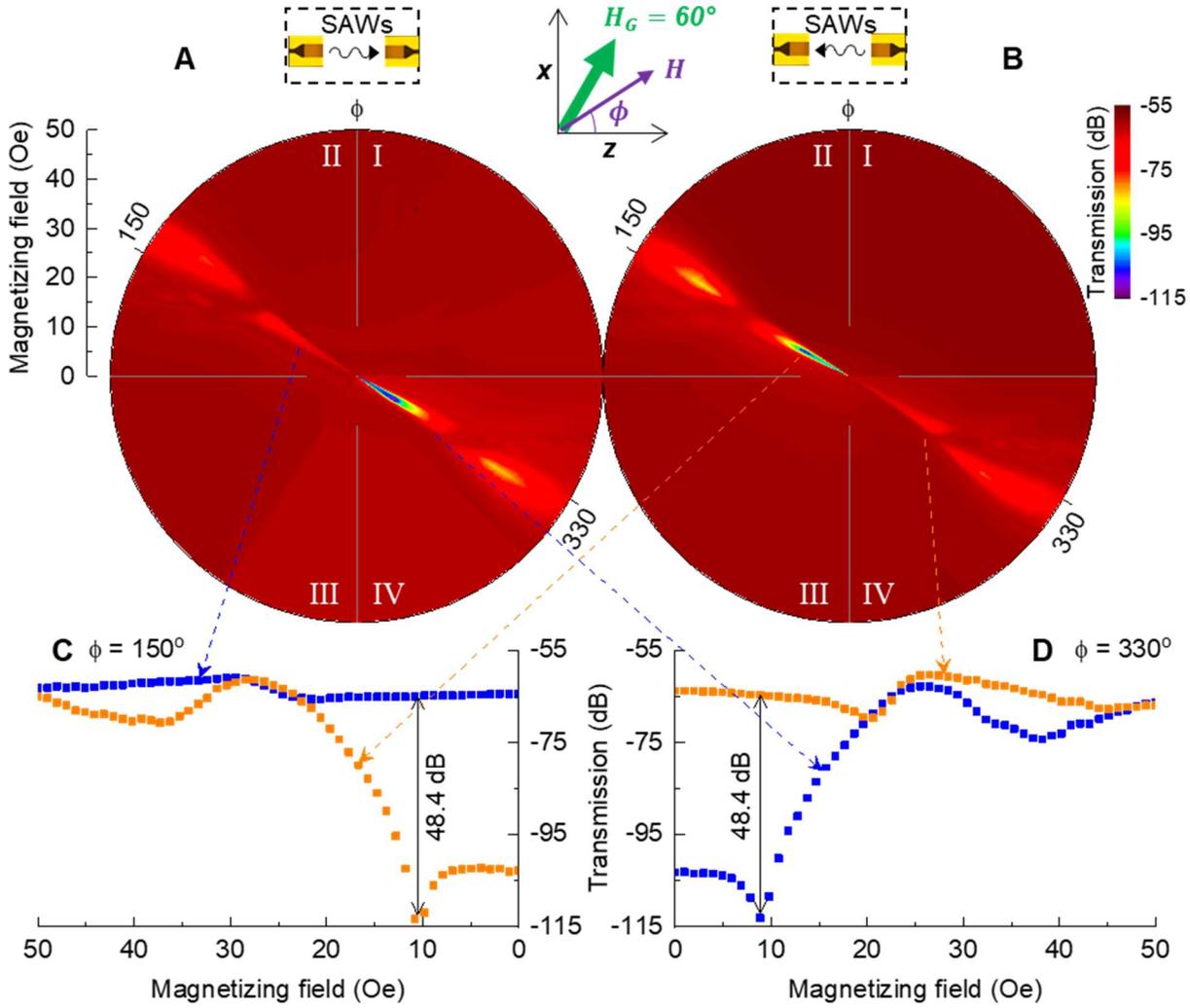

**Fig. 2. Nonreciprocal SAW transmission in FeGaB/Al$_2$O$_3$/FeGaB multilayer stack with growth field $H_G$ at 60°.** ADFMR plot with 1435-MHz SAWs traveling in (**A**) +z-direction (forward) and (**B**) -z-direction (reverse). Absorption (blue on color scale) occurs at ADFMR resonance, which we observe in directions perpendicular to the growth field. (**C**) Field sweeps at $\phi = 150°$ for forward (blue) and reverse (orange) SAW propagation. (**D**) Field sweeps at $\phi = 330°$ for forward (blue) and reverse (orange) SAW propagation. The difference between forward and reverse sweeps at a common static field condition is the isolation.

In Fig. 2, red indicates maximum transmission (far from magnetic resonance) of approximately -55 dB. This is the insertion loss of the device and there are several known reasons for high insertion loss. Insertion loss increases with frequency due to interactions with thermally excited elastic waves and energy lost to air adjacent to the surface (air loading) [17]. In addition, operating the device at higher order harmonics and impedance mismatch also contribute to higher insertion loss, but these can be minimized through careful engineering of the IDTs which is outside the scope of the current work.

Resonant absorption of SAWs by spin waves in the magnetic material appear as blue on the color scale and we observe none of this interaction along the growth field axis 60° (quadrant I) and 240° (quadrant III). Interactions only exist in the perpendicular directions 150° (quadrant II) and 330° (quadrant IV). Minimum transmission was -115 dB, which occurred at $H = 9$ Oe, $\phi = 150°$ for forward propagating SAWs, and $H = 11$ Oe, $\phi = 330°$ for reverse propagating SAWs. This minimum transmission value is 60 dB lower than the insertion loss (99.9999% power absorption) which is remarkable in its own right as the highest reported ADFMR-related absorption value to date. These conditions of maximum acoustic-spin interaction are highlighted by the line cuts in Fig. 2C (orange) and Fig. 2D (blue). Reversing the SAW propagation direction under the same field conditions results in the other line cuts in Fig. 2C (blue) and Fig. 2D (orange), which are nearly flat in comparison indicating almost zero interaction between the SAWs and the magnetic layer.

The isolation (i.e. the difference between forward and reverse transmission) measured at both resonant angles is 48.4 dB. This represents significantly higher isolation performance compared to state of the art commercial isolator devices. There is a second resonance at higher field ($H = 35$ Oe, $\phi = 150°$) which also exhibits significant nonreciprocity, albeit with much lower isolation than the low-field resonance. Results were qualitatively similar with a driving SAW frequency of 863 MHz, but with a lower isolation of about 10 dB (See Supplementary Information).

When $H_G = 0$ or $90°$, the nonreciprocity minimizes or completely disappears. Here, absorption occurs at small angles from the $z$-axis, but along the $z$-axis ADFMR is prohibited because there exists no $x$-component for the microwave driving field produced by the Rayleigh waves, which is a necessary condition for FMR.

From the line cuts in Fig. 3C and Fig. 3D we do observe small nonreciprocity. Small absorption of SAWs occurred even if the anisotropy axis was oriented non-optimally. The absorption happens because the magnetic state of the bilayers became canted under the external magnetic field and some portion the magnetoelastic driving field can interact with the magnetic material. Nonetheless, this is negligible compared to the giant nonreciprocity shown in Fig. 2. Fig. 2 and Fig. 3 are shown on the same scale for ease of comparison.

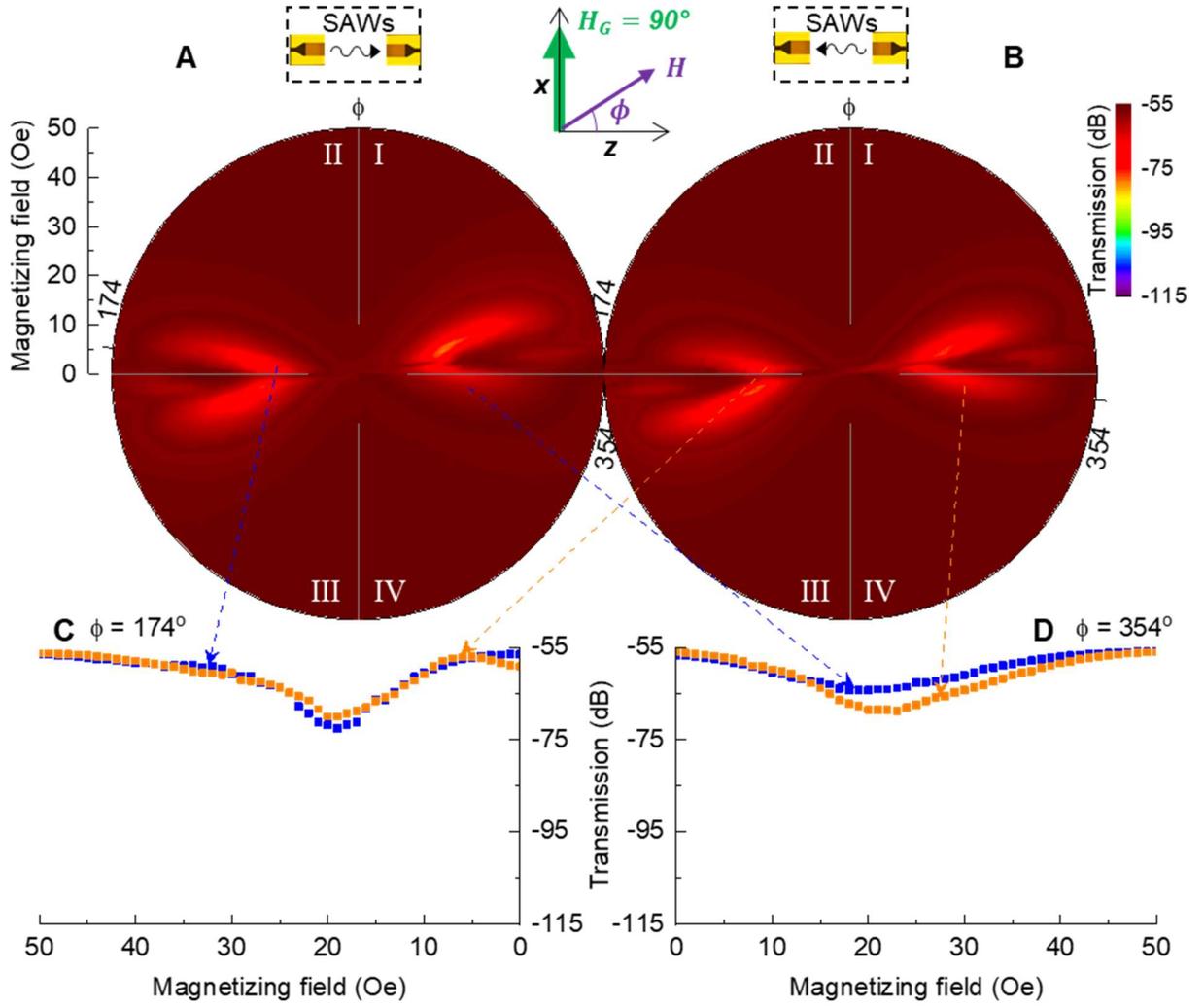

**Fig. 3. SAW transmission in FeGaB/Al$_2$O$_3$/FeGaB multilayer stack with growth field $H_G$ at 90°.** ADFMR plot with 1435-MHz SAWs traveling in (**A**) +z-direction (forward) and (**B**) -z-direction (reverse). Absorption (blue on color scale) occurs at ADFMR resonance, close to 0 and 180° in this case. (**C**) Field sweeps at $\phi = 174°$ for forward (blue) and reverse (orange) SAW propagation. (**D**) Field sweeps at $\phi = 354°$ for forward (blue) and reverse (orange) SAW propagation.

## Discussion

### Spin wave spectra in bilayer heterostructures

A general theory of spin-wave propagation in antiparallel magnetic layers has been developed in [18]. Here we build on that derivation by developing the theory of magnetoelastic interactions. We consider two parallel ferromagnetic film planes. For simplicity we assume the films are composed of identical magnetic material and have identical thickness $L$. The layers are separated by a nonmagnetic spacer with thickness $d_s$. Also we assume that the layers are sufficiently thin ($L \ll \lambda$) with respect to spin-wave wavelength $\lambda$.

Magnetic dynamics in each layer are governed by Landau-Lifshitz equations:

$$\frac{dM_i(t)}{dt} = \gamma B_i^{\text{eff}}(t, M_1, M_2) \times M_i(t), \tag{1}$$

where $i = 1,2$ is the layer index, $\gamma/(2\pi) \approx 28$ GHz/T is the gyromagnetic ratio, $M_i$ is the magnetization in each layer and $B_i^{eff}$ is the effective magnetic field acting on the $i$-th layer. For the sake of simplicity in the following derivation we assume that the magnetization is uniform across a layer, but can be different in each layer. The full derivation is discussed at length in [18].

Here we are interested in the small amplitude dynamics and we can decompose the magnetization into static and dynamic parts as $M_i(t) = M_s(\mu_i + s_i e^{-i\omega t})$, where $|\mu| = 1$ is the vector pointing in the direction of the equilibrium magnetization, $s_i$ is the spin-excitation vector, and $\omega$ is the angular frequency of the spin-wave. By definition these vectors are orthogonal to each other: $\mu_i \cdot s_i = 0$. This expansion allows us to linearize (1) using the standard procedure discussed in [19, 20].

Assuming the lateral dimensions of the films are much larger than the distance between them, we can neglect the static part of the interaction between layers. Under this approximation we can find the equilibrium magnetization and the internal magnetic field $B_i$ as:

$$B_i \mu_i = B^{ext} - \mu_0 M_s (\widehat{N}_0 + \widehat{K}) \cdot \mu_i \qquad (2)$$

where $\widehat{N}_0$ and $\widehat{K}$ are the tensors of static demagnetization and uniaxial anisotropy.

The dynamic part of (1) can be written as:

$$-i\omega_k s_i = \mu_i \times [\widehat{\Omega}_k \cdot s_i + \omega_M \widehat{R}_k^{i,j} \cdot s_j] \qquad (3)$$

where $\widehat{\Omega}_k = [\gamma B_i \widehat{I} + \omega_M(\widehat{N}_k + \widehat{K} + \lambda_{ex}^2 k^2 \widehat{I})]$ is the dynamic self-interaction tensor [20, 21], $\widehat{N}_k$ is the dynamic self-demagnetization tensor [19], $\widehat{I}$ is an identity matrix, $\widehat{R}_k^{i,j}$ is the mutual cross-demagnetization tensor which defines the dynamic interaction between the films, $\lambda_{ex}$ is the inhomogeneous exchange length, $j \neq i$, $\omega_M = \gamma \mu_0 M_s$, and $k$ is the wave vector of the spin wave. Wave vector dependence is implied for $s_i$, although the index is dropped.

Equation (3) can be rewritten in a more compact form of a standard eigenvalue problem [20]:

$$-i\omega_k \begin{pmatrix} \widehat{J}_1 & 0 \\ 0 & \widehat{J}_2 \end{pmatrix} \cdot \begin{pmatrix} s_1 \\ s_2 \end{pmatrix} = \widehat{\mathcal{P}} \cdot \begin{pmatrix} \widehat{\Omega}_1 & \omega_M \widehat{R}_k^{1,2} \\ \omega_M \widehat{R}_k^{2,1} & \widehat{\Omega}_2 \end{pmatrix} \cdot \widehat{\mathcal{P}} \cdot \begin{pmatrix} s_1 \\ s_2 \end{pmatrix}, \qquad (4)$$

where $\omega_M = \gamma \mu_0 M_s$, $\widehat{J}_i = \hat{e} \cdot \mu_i$, $\hat{e}$ is the Levi-Civita tensor, and

$$\widehat{\mathcal{P}} = -\begin{pmatrix} \widehat{J}_1 \cdot \widehat{J}_1 & 0 \\ 0 & \widehat{J}_2 \cdot \widehat{J}_2 \end{pmatrix}.$$

The solution to (4) is obtained with standard numerical methods obtaining values of $\omega$ and $s_i$ as a function of $k$.

To obtain the explicit form of tensors $\widehat{N}_k$ and $\widehat{R}_k^{i,j}$ we fix the coordinate system as $k = k\mathbb{x}$ and $\mathbb{y}$ is the normal to the film surface. In this coordinate system the self-demagnetization tensor [19] is $\widehat{N}_k = p\mathbb{x} \otimes \mathbb{x} + (1-p)\mathbb{y} \otimes \mathbb{y}$ with $p = (-1 + |k|L + e^{-|k|L})/(|k|L)$.

The mutual demagnetization tensor is $\widehat{R}_k^{1,2} = g(\mathbb{x} \otimes \mathbb{x} - \mathbb{y} \otimes \mathbb{y}) + ig\,\text{sign}(k)(\mathbb{y} \otimes \mathbb{x} + \mathbb{x} \otimes \mathbb{y})$, where

$$g = \int_0^L \int_{L+d_s}^{2L+d_s} \frac{1}{2L} e^{-|k||y-y'|} dy dy' = \frac{e^{-|k|(d_s+2L)}(-1+e^{|k|L})^2}{2|k|L}. \tag{5}$$

Note that $\widehat{R}_k^{1,2} = (\widehat{R}_k^{2,1})^\dagger \neq (\widehat{R}_{-k}^{1,2})$, which constitutes the necessary condition for spin wave nonreciprocity [21, 22, 23, 3, 24].

**Magnetoelastic coupling**

The magnetoelastic interaction couples spin waves in the magnetic film and SAWs in the substrate. This coupling leads to a modification in the dispersion characteristics of the SAWs, ultimately changing the propagation behavior. Here we are interested in the modification of losses incurred by SAWs traveling in opposite directions.

A general theory of SAW/spin wave interactions has been developed by Verba et al. [2, 4]. Here we employ several approximations to make the analytical calculations manageable. First, we consider the magnetic layer acoustically identical to the substrate material, i.e. we do not take into account the mass loading effect. In general, mass loading is important for SAW IDT matching, however, as the mass loading is a purely mechanical effect it does not contribute to nonreciprocity. Second, we assume that the magnetoelastic coupling energy is much smaller than other interaction energies in the system, which is practically always true for ferromagnets with strong magnetostriction.

In magnetostrictive materials, acoustic and magnetic systems are coupled via magnetoelastic interaction with characteristic energy density:

$$W^{me}(r) = \frac{b_{ijkl} u_{ij}(r) M_i(r) M_j(r)}{M_s^2}, \tag{6}$$

where $b_{ijkl}$ is the magnetoelastic tensor and $\hat{u}$ is the mechanical strain-tensor of the SAWs [25]. Here we assume the magnetoelastic coupling to be isotropic and uniform across the ferromagnetic sample $b_{ijkl} = b\delta_{ij}\delta_{kl} + sym$. Note that the energy density itself does depend on the direction and position in the sample.

The magnetoelastic interaction entangles SAW and spin wave modes. In the weakly coupled oscillator model, the spectrum of magnetoelastic waves can be found as described in [26] as

$$\omega_k^{me} = \frac{1}{2}(\omega_k^a + \omega_k^{sw}) \pm \sqrt{\frac{(\omega_k^a - \omega_k^{sw})^2}{4} + |\kappa|^2} \tag{7}$$

where $\omega_k^a$ is the dispersion of the acoustic wave, $\omega_k^{sw}$ is the dispersion of the spin wave and $\kappa$ is the intermode coupling coefficient [4]. The coupling coefficient can be found as an overlap integral of the SAW and spin wave mode

$$\kappa = \frac{2b}{\sqrt{A}\sqrt{Q}} \int_{-\infty}^{0} \mu(y) \cdot \hat{u}^{\dagger}(y) \cdot s(y) dy \tag{8}$$

where † stands for Hermitian conjugation and the coefficients $\sqrt{A}$ and $\sqrt{Q}$ are the normalizing constants calculated below [4]. In general, it is difficult to obtain useful explicit expressions for the coupling coefficient for an arbitrary configuration of the external magnetic field, magnetic anisotropy, and SAW propagation direction. Some particular cases for a bilayer magnetic film have been considered in [2]. Here we use the closed form (8) and evaluate the integral numerically. The mechanical strain-tensor for a SAW propagating in the $z$-direction can be obtained as [25]:

$$u_{zz} = e^{ik_R z} k_R^{-1} (k_R e^{\kappa_l y} + a\kappa_t e^{\kappa_t y}) \tag{9}$$

$$u_{yy} = -ie^{ik_R z} k_R^{-2} (\kappa_l^2 e^{\kappa_l y} + ak_R \kappa_t e^{\kappa_t y}) \tag{10}$$

$$u_{yz} = u_{zy} = \frac{1}{2} e^{ik_R z} k_R^{-2} (2k_R \kappa_l e^{\kappa_l y} + a(\kappa_t^2 + k_R^2) e^{\kappa_t y}) \tag{11}$$

$$Q = |k_R^5| \zeta \rho c_t \left( 4ak_R + \frac{(k_R^2 + \kappa_l)^2}{2\kappa_l} + \frac{a^2(k_R^2 + \kappa_t^2)}{\kappa_t} \right), \tag{12}$$

where $\zeta$ is the Rayleigh coefficient, $c_t$ and $c_l$ are transverse and longitudinal velocities, $k_R = \zeta \omega/c_t$ is the Rayleigh wavenumber, $\rho$ is the density, $a = 1/2k_R/|k_R|(2-\zeta^2)/\sqrt{1-\zeta^2}$, $\kappa_t = |k_R|\sqrt{1-\zeta^2}$ and $\kappa_t = |k_R|\sqrt{1-c_t/c_l\zeta^2}$.

The magnetic part of (8) can be obtained in the approximation of a uniform magnetization distribution across the film thickness. The static part is obtained by solving (2) and using the following function in (8):

$$\mu(y) = \mu_1 \Pi_1(y) + \mu_2 \Pi_2(y), \tag{13}$$

where $\Pi_1 = \theta(y+L) - \theta(y), \Pi_2 = \theta(y+2L+d_s) - \theta(y+L+d_s)$, and $\theta(y)$ is the Heaviside step function. The dynamic part can be constructed analogously by solving (4) and using the following function: $s(y) = s_1 \Pi_1(y) + s_2 \Pi_2(y)$. The magnetic normalizing constant $A$ can be found from the expression [4, 21].

$$A = i\frac{2LM_s}{\gamma} \sum_{j=1,2} \mu_i \cdot (s_j^{\dagger} \times s_j). \tag{14}$$

Note that $A$ and $Q$ have the same dimensionality of the action linear density.

**Damping of magnetoelastic waves**

The operating principle of the magnetic SAW isolator is the direction selectivity of the SAW damping. This damping is due to the loss of energy in the magnetic system which is usually several orders of magnitude larger than intrinsic SAW losses.

To take into account magnetic losses we substitute the magnetic eigenfrequency as $\omega_k \to \omega_k + i\Gamma_k$ where the decay rate is calculated as

$$\Gamma = \alpha_G \omega_k \frac{2LM_s}{\gamma A} \sum_{i=1,2} s_i^{\dagger} \cdot s^{\dagger}. \tag{15}$$

Where $\alpha_G$ is the Gilbert damping constant. The linear magnetic losses of the SAWs can be approximately calculated from (7):

$$\text{Loss} \approx 4.34\, \text{Im}(\omega^{me})/(c_R)\ [\text{dB/m}], \tag{16}$$

where $c_R = \zeta c_t$ is the Rayleigh SAW velocity.

Collectively based on the above calculation, we can calculate the damping caused by magnetoelastic coupling in SAWs. For our calculations we take the following parameters for the magnetic material: $\mu_0 M_s = 1.3$ T, $\gamma = 28$ GHz/T, $\alpha_G = 2\times10^{-2}$, $b = 9.38$ MJ/m³, $L = 20$ nm, $d_s = 5$ nm, $b_a = 1.8$ mT, $\lambda_{ex} = 4.7$ nm. The elastic properties for LiNbO3 are: $c_l = 2.8$ km/s, $c_t = 3.85$ km/s, $\rho = 4650$ kg/m³.

First let us consider the case when the anisotropy axis is oriented at $\theta_m = 60°$ to the direction of SAW propagation. The external magnetic field is applied perpendicular to the anisotropy axis. For SAW excitation frequency $\omega/(2\pi) = 1435$ MHz, the dependence of the losses is plotted in Fig. 4A. Note the large peak situated around the value of 10 Oe for the applied magnetic field. The maximum absorption of the SAWs happens when the SAW dispersion curve crosses the dispersion curve of spin waves. To illustrate this point we plot the SAW and spin wave dispersion for the applied magnetic field of 11 Oe. Note that two dispersion curves for the spin waves and SAWs cross at 1435 MHz for the negative values of wavenumber (backward propagation), while they stay apart for forward propagation. The consequence of this crossing is the enhanced absorption of backward propagating waves.

The results in Fig. 4A show that theoretically at 10 Oe applied magnetic field, the forward loss is around 0.4 dB/mm (indicating near full transmission) whereas the backward loss is about 7.5 dB/mm. Taking the difference between forward and backward loss, 7.1 dB/mm, and multiplying by the 2.2-mm length of the multilayer stack, the calculated isolation performance is around 15.6 dB. When the growth field is applied at 90° (Fig. 4B), the calculated isolation performance is only about 0.2 dB, nearly negligible in comparison with the 60° case.

Experimentally, in our measurements for $\theta_m = 60°$ the difference between forward and backward propagation is the isolation performance of the device, which is remarkably high isolation of 48.4 dB. We suspect that since the resonance fields are very low in these multilayer stacks, assumptions of the film properties will have strong influence on the theoretical prediction of the device performance, particularly the assumption of gyromagnetic ratio and the Gilbert damping constant. This may explain the difference between the theoretical value and our experimental result.

Nevertheless these results are particularly interesting for the development of next generation frequency-agile tunable isolator and circulator devices because of the high isolation performance. One obvious concern to address in the near term is the high insertion loss, which can be mitigated by advanced SAW IDT design techniques taking care to impedance match at the desired operating frequency.

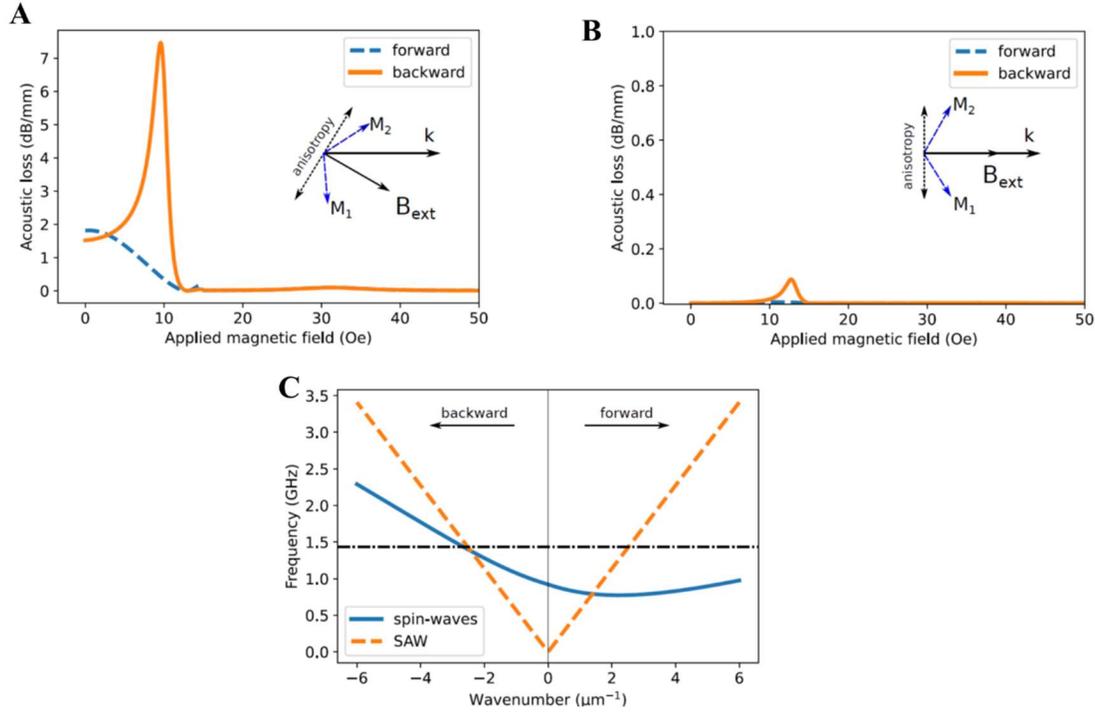

**Fig. 4. Theoretical behavior of the nonreciprocal ADFMR device concept.** (**A**) Linear loss for the magnetoelastic wave as a function of the applied magnetic field calculated for forward and backward propagation for $H_G = 60°$. The inset shows the configuration of the SAW propagation, applied magnetic field and crystalline anisotropy. (**B**) Same as (**A**) for $H_G = 90°$. (**C**) Dispersion of spin-waves in a bilayer and SAWs in the substrate for applied magnetic field of 11.5 Oe.

**Materials and Methods**

In this work, we use split-finger IDT design, generating Rayleigh waves using single crystal *y*-cut LiNbO$_3$ substrate. Favorable SAW propagation is along the *z*-axis between two pairs of IDTs for the delay line filter geometry as shown in the schematic representation of the device in Fig. 1. Split-finger design minimizes the destructive interference caused by reflection from the IDTs and thereby allows the device operation at higher odd harmonics of the fundamental frequency. The nominal designed fundamental frequency $f_1$ is around 291 MHz, however, most of the reported measurements are at higher harmonic $f_3$, $f_5$, and $f_7$ = 873, 1455 and 2037 MHz, respectively. IDTs have 60 finger pairs with the minimum electrode separation $\lambda/8$ = 1.5 µm. The delay line spacing between two IDT pairs is 3 mm. IDT patterning for metal-liftoff was completed using negative tone lift-off photoresist NR9-1000Py and a Karl Suss MA6 mask aligner contact lithography system. The Al electrode thickness is 70 nm, deposited using e-beam evaporation. Details on the SAW device design, fabrication and its impact on ADFMR performance are discussed in our prior work [15].

In the spacing between the IDTs, a multilayer thin-film stack of FeGaB/Al$_2$O$_3$/FeGaB is deposited via sputtering and lithographically patterned [27] with width 500 µm along *x* and length 2200 µm along *z*. The film stack is deposited in the presence of external magnetic field. A schematic figure of in-situ magnetic field orientation with respect to the SAW *k*-vector is shown in the Supplemental Information section.

For measurements, a signal generator delivered pulsed RF with 20 dBm power to the ADFMR device, and the amplified output was measured using a spectrum analyzer. Time-gating was used to isolate the signal transmitted via SAWs, which is delayed by about 1 μs compared to the EM radiative signal because of the slower velocity of SAWs. A vector electromagnet was used to sweep the angle $\phi$ and magnitude $H$ of the magnetizing field. The magnitude was swept from high (50 Oe) to low (0 Oe) to ensure consistency in hysteretic behavior. To measure the nonreciprocal transmission behavior of the device, the generator and analyzer were interchanged between input and output ports in two separate measurement sweeps. A vector network analyzer was used to calibrate the transmission values.

## Acknowledgments

This work was supported by the Air Force Office of Scientific Research (AFOSR) Award No. FA955020RXCOR074.

# Supplementary Information for

# Giant Nonreciprocity of Surface Acoustic Waves enabled by the Magnetoelastic Interaction

**Authors**

Piyush J. Shah[1], Derek A. Bas[1], Ivan Lisenkov[2], Alexei Matyushov[3,4], Nianxiang Sun[3], Michael R. Page[1*]

**Affiliations**

[1] Materials and Manufacturing Directorate, Air Force Research Laboratory, Wright-Patterson Air Force Base, Ohio 45433, USA
[2] Independent Researcher, Newton Upper Falls, MA 02464
[3] Department of Electrical and Computer Engineering, Northeastern University, Boston, MA 02115, USA
[4] Department of Physics, Northeastern University, Boston, MA 02115, USA

*Email: michael.page.16@us.af.mil

**Supplemental figures**

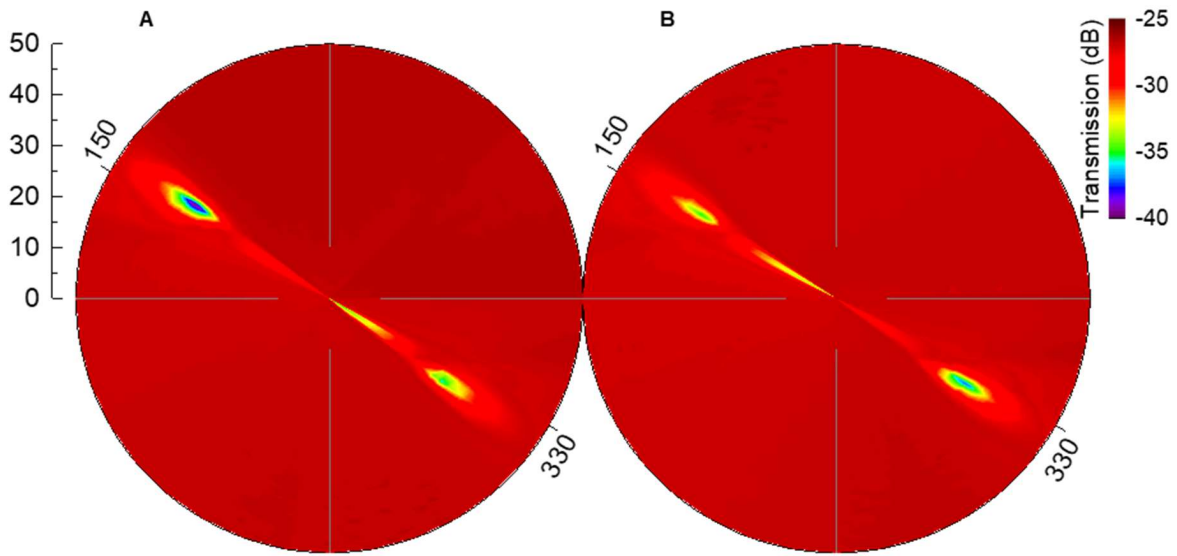

**Fig. S5. Nonreciprocal SAW transmission in FeGaB/Al$_2$O$_3$/FeGaB multilayer stack with growth field at 60°.** (**A**) ADFMR plot with 863-MHz SAWs traveling in +z-direction (forward). (**B**) ADFMR plot with 863-MHz SAWs traveling in -z-direction (reverse).

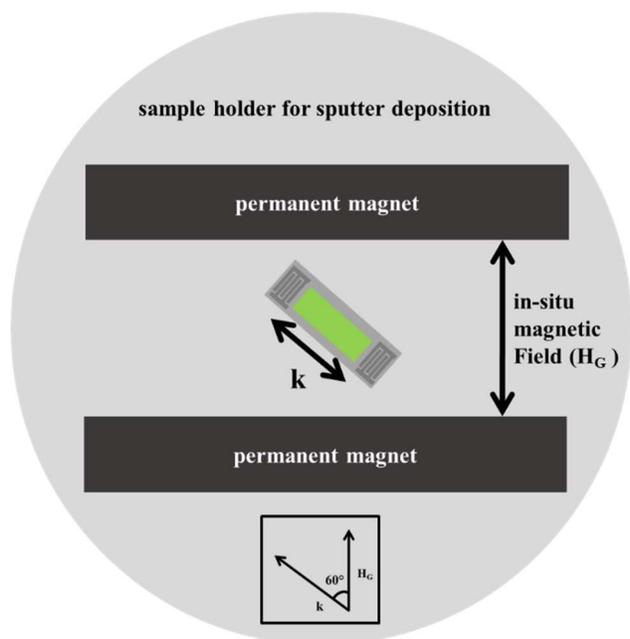

**Fig. S6. Schematic diagram of the sample holder for sputter deposition with in-situ magnetic field.** Sample is place in between two bar permanent magnets. Inset shows the angle relationship between the applied growth fields with respect to the SAW k vector direction. Samples are physically placed in between the permanent bar magnets and measured field along the center line is about 200 Oe.

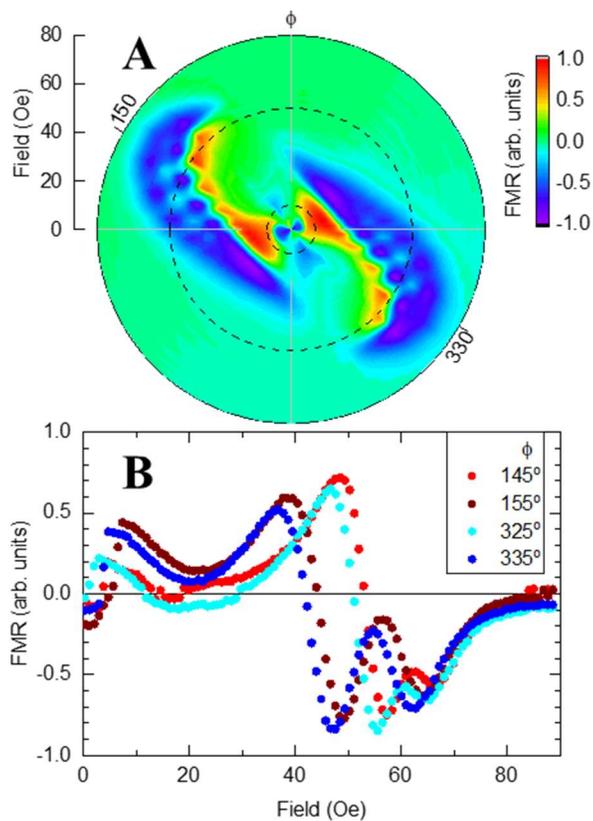

**Fig. S7. Ferromagnetic resonance at 2 GHz.** (**A**) Angle-dependence of FMR showing extreme anisotropy defined by the growth field at 60°. (**B**) Line cuts at angles near the optimal isolator conditions. Multiple resonances occur near 5 and 45 Oe, similar to that seen in ADFMR. In FMR the uniform mode is excited, a notable difference from ADFMR in which the spin waves are traveling.